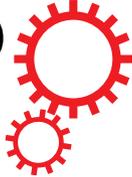

**OPEN** Electric field control of magnon-induced magnetization dynamics in multiferroics

Vetle Risinggård*, Iryna Kulagina* & Jacob Linder

We consider theoretically the effect of an inhomogeneous magnetoelectric coupling on the magnon-induced dynamics of a ferromagnet. The magnon-mediated magnetoelectric torque affects both the homogeneous magnetization and magnon-driven domain wall motion. In the domains, we predict a reorientation of the magnetization, controllable by the applied electric field, which is almost an order of magnitude larger than that observed in other physical systems via the same mechanism. The applied electric field can also be used to tune the domain wall speed and direction of motion in a linear fashion, producing domain wall velocities several times the zero field velocity. These results show that multiferroic systems offer a promising arena to achieve low-dissipation magnetization rotation and domain wall motion by exciting spin-waves.

Multiferroic materials are materials that exhibit simultaneously two or more ferroic order parameters (ferroelectric, (anti)ferromagnetic, ferroelastic and ferrotoroidic order)[1]. The term *multiferroics* has also been used more generally to refer to materials that exhibit a magnetoelectric coupling—that is, materials with a free energy functional that contains terms dependent on both the magnetization and the polarization[2–4].

If the cross-terms in the free energy only depend on the magnetization itself—and not its derivatives—the material is said to exhibit the homogeneous magnetoelectric effect. The first detailed investigation of this effect was made by Dzyaloshinskii[5] for $Cr_2O_3$. An essential aspect of the homogeneous magnetoelectric effect is that an effect that is first order in both the polarization and the magnetization (linear magnetoelectric effect) cannot occur in materials where the free energy has time-reversal and spatial inversion symmetry[6].

If the cross-terms in the free energy depend on the gradient of the magnetization, the material is said to exhibit the inhomogeneous magnetoelectric effect, or the flexoelectric effect. The existence of this effect was first pointed out by Bar'yakhtar *et al.*[7]. The inhomogeneous magnetoelectric effect can be present even in inversion symmetric systems. The study of the inhomogeneous magnetoelectric effect has seen a revival after the work of Mostovoy[8] on the rare earth manganites. Several authors have contributed to an extension of these results to the dynamic regime[9–16].

In this work, we determine how the magnetization dynamics of multiferroic materials are influenced by injecting spin-waves. Magnon spintronics is an emerging subfield of the field of spintronics which applies spin-waves for information transport and processing[17,18]. By carrying spin currents using magnons rather than electrons, large scale charge transport and the associated Joule heating is avoided. Magnons can propagate over centimeter distances in low-damping magnetic insulators[19], while spin-currents carried by electrons are limited by the spin-diffusion length, which is on the order of microns. Magnons also offer exciting possibilities exploiting wave-based and nonlinear phenomena, such as the majority gate[20–22] and parallel computing[23].

One of the principal advantages of magnon spintronics is the wide variety of available magnetic materials and interactions, and the large number of other magnetic excitations such as domain walls, vortices and skyrmions. For instance, there are obvious opportunities for creating nonvolatile memories based on magnons interacting with domain walls[24,25] or skyrmions[26,27]. The properties of such systems can be tuned by exploiting higher order magnetic interactions such as the Dzyaloshinskii–Moriya interaction (DMI)[28,29], not to mention magnon band structure engineering in magnonic crystals[18].

We will first consider the effect of an inhomogeneous magnetoelectric coupling on magnetization dynamics induced by magnon injection. We demonstrate analytically and numerically that this term produces a reorientation of the time-averaged magnetization, see Fig. 1. A similar reorientation has been identified previously by

NTNU, Norwegian University of Science and Technology, Department of Physics, N-7491 Trondheim, Norway. *These authors contributed equally to this work. Correspondence and requests for materials should be addressed to V.R. (email: vetle.m.risinggard@ntnu.no)

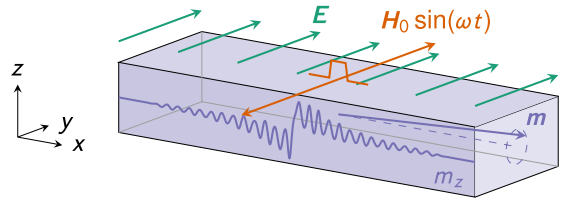

**Figure 1. Magnetization reorientation by the magnon-mediated magnetoelectric torque.** When an electric field is applied to a homogeneously magnetized sample, nothing happens. When a magnetic inhomogeneity is introduced in the form of spin-waves, the interaction between the applied electric field and the induced electric polarization produces a shift in the time-averaged magnetization that is linear in the applied electric field.

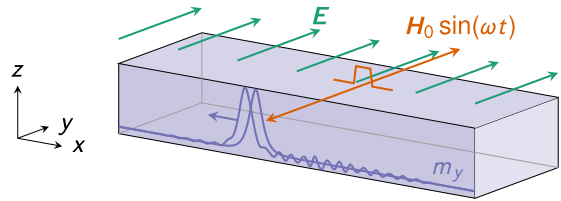

**Figure 2. Controlling the domain wall velocity by the magnon-mediated magnetoelectric torque.** In the absence of an applied electric field, the magnetic domain wall travels towards the spin-wave source. The magnitude of the velocity is determined by the spin-wave excitation amplitude, the distance from the source and the spin-wave frequency. By application of an electric field the domain wall can be made to stop and change direction of motion. The velocity of the domain wall is linear in the electric field.

Manchon *et al.*[30] for a Dzyaloshinskii–Moriya ferromagnet and by Linder[31] for a topological insulator–ferromagnet heterostructure. Unlike these previous results, the magnitude and direction of the reorientation of the magnetization that is due to the inhomogeneous magnetoelectric effect is controllable by the applied electric field; not fixed by the material constants. Moreover, the effect is quantitatively much larger than in the Dzyaloshinskii–Moriya case.

Second, we determine how the inhomogeneous magnetoelectric effect alters magnetic domain wall motion. The case of magnetic field-driven domain wall motion has been considered previously by Chen *et al.*[32], whereas Dzyaloshinskii[33] and Logginov *et al.*[34,35] have considered domain wall motion driven by an inhomogeneous electric field. Magnon-induced domain wall motion in ordinary ferromagnets was first considered by Mikhailov and Yaremchuk[36], and more recently by Yan *et al.*[24] who showed that conservation of angular momentum will drive the domain wall towards the spin-wave source. In response to puzzling numerical results showing domain wall motion away from the spin-wave source, Yan *et al.*[37] have developed a theory of linear momentum transfer from spin-waves to domain walls. Linear momentum transfer has later been used to explain the dependence of the direction of domain wall motion in Dzyaloshinskii–Moriya ferromagnets on the sign of the material dependent DMI constant[38]. Interestingly, we find that for magnon-induced domain wall motion the inhomogeneous magnetoelectric effect enables electric field control of both the sign and magnitude of the domain wall velocity, see Fig. 2. The domain wall velocity scales linearly with the applied electric field.

## Magnon-induced dynamics in homogeneously magnetized multiferroics

As shown in Fig. 1, we consider a ferromagnetic wire that exhibits the inhomogeneous magnetoelectric effect. In the continuum limit, the magnetization can be described by a vector field $m(r, t)$. The free energy of this system can be written as

$$F[m] = \frac{A}{m^2}\int\left[(\nabla m_x)^2 + (\nabla m_y)^2 + (\nabla m_z)^2\right]d\mathbf{r} - m^{-2}\int(Km_x^2 - K_\perp m_z^2)\,d\mathbf{r} - \int(\mathbf{E}\cdot\mathbf{P})\,d\mathbf{r}. \quad (1)$$

Here, $A$ is the exchange stiffness, $m$ is the saturation magnetization, $K$ is the easy axis anisotropy constant and $K_\perp$ is the hard axis anisotropy constant. The easy axis of magnetization is taken to be along the length of the wire and is dominated by shape anisotropy. If the wire cross-section is not circular, the shape anisotropy will also contribute a perpendicular hard axis anisotropy. The effective constant $K_\perp$ can also contain contributions from magnetocrystalline anisotropy. The final term is the inhomogeneous magnetoelectric interaction. $\mathbf{E}$ is the applied electric field and $\mathbf{P}$ is the induced electric polarization[8],

$$\mathbf{P} = \gamma_0[\mathbf{m}(\nabla\cdot\mathbf{m}) - (\mathbf{m}\cdot\nabla)\mathbf{m}]. \quad (2)$$

www.nature.com/scientificreports/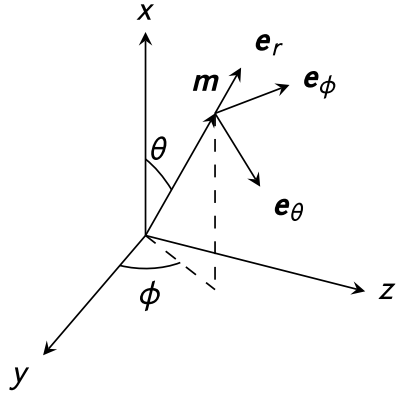

**Figure 3. For the spin-wave perturbation theory the magnetization is most conveniently written out in a spherical coordinate system.** When the magnetization is homogeneous, $\theta$ and $\phi$ are assumed to be independent of position and time. In the case of a domain wall our objective is to consider the *dynamic* response of a *nonuniform* magnetization texture. Consequently, $\theta$ and $\phi$ must be position- and time-dependent.

$\gamma_0$ is the inhomogeneous magnetoelectric coupling constant.

We describe the time dynamics of the magnetization using the Landau–Lifshitz–Gilbert (LLG) equation[39–41],

$$\partial_t \boldsymbol{m} = \gamma \boldsymbol{m} \times \boldsymbol{H} - \frac{\alpha}{m} \boldsymbol{m} \times \partial_t \boldsymbol{m}, \tag{3}$$

where $\boldsymbol{H}(\boldsymbol{r}, t)$ is the effective magnetic field acting on the magnetization, $\gamma$ is the gyromagnetic ratio and $\alpha$ is the Gilbert damping parameter. We assume that $\gamma < 0$ and $\alpha < 0$. The effective magnetic field is given by the functional derivative of the free energy with respect to the magnetization, $\boldsymbol{H} = -\delta F/\delta \boldsymbol{m}$. This gives rise to the effective field

$$\boldsymbol{H} = \frac{2A}{m^2}\nabla^2 \boldsymbol{m} + \frac{2K}{m^2} m_x \boldsymbol{e}_x - \frac{2K_\perp}{m^2} m_z \boldsymbol{e}_z + 2\gamma_0[(\nabla \cdot \boldsymbol{m})\boldsymbol{E} - \nabla(\boldsymbol{E} \cdot \boldsymbol{m})], \tag{4}$$

where we have assumed that the electric field is independent of position.

**Perturbation theory.** We calculate the magnon-induced dynamics perturbatively by considering deviations from a homogeneous magnetization parametrized in the small excitation parameter $h$. For this purpose the magnetization is most conveniently written out in a spherical frame as shown in Fig. 3. To second order in $h$ the magnetization is[42]

$$\begin{aligned}
\boldsymbol{m}(\boldsymbol{r}, t) &= \left(m - h^2\left[s_\theta^{(2)}(\boldsymbol{r}, t) + s_\phi^{(2)}(\boldsymbol{r}, t)\right]\right)\boldsymbol{e}_r + \left[h s_\theta^{(1)}(\boldsymbol{r}, t) + h^2 \Delta m_\theta(\boldsymbol{r})\right]\boldsymbol{e}_\theta \\
&\quad + \left[h s_\phi^{(1)}(\boldsymbol{r}, t) + h^2 \Delta m_\phi(\boldsymbol{r})\right]\boldsymbol{e}_\phi \\
&= \left(m - \frac{h^2}{2m}[s_\theta^2(\boldsymbol{r}, t) + s_\phi^2(\boldsymbol{r}, t)]\right)\boldsymbol{e}_r + [h s_\theta(\boldsymbol{r}, t) + h^2 \Delta m_\theta(\boldsymbol{r})]\boldsymbol{e}_\theta \\
&\quad + \left[h s_\phi(\boldsymbol{r}, t) + h^2 \Delta m_\phi(\boldsymbol{r})\right]\boldsymbol{e}_\phi.
\end{aligned} \tag{5}$$

$s_{\theta/\phi}^{(1)/(2)}(\boldsymbol{r}, t)$ is the first (second) order spin-wave excitations in the $\theta$- ($\phi$)-direction. For small $h$ the linear terms dominate, reproducing the usual spin-wave expansion[24]. The second order excitations take into account the fact that the magnetization along the $r$-direction is reduced for stronger excitation[30]. $\Delta m_{\theta/\phi}(\boldsymbol{r})$ correspond to a steady-state reorientation of the magnetization that might be induced by the spin-waves. Since spin-waves in an ordinary ferromagnet do not induce such deviations, these deviations must be of second order in $h$ or higher[30]. The second equality follows from the micromagnetic normalization criterion $\boldsymbol{m} \cdot \boldsymbol{m} = m^2$ and is accurate to second order in $h$. For ease of notation, we have written $s_{\theta/\phi}^{(1)}(\boldsymbol{r}, t) = s_{\theta/\phi}(\boldsymbol{r}, t)$.

*Static magnetization.* For simplicity, we assume that our system is one-dimensional so that $\nabla = \partial_x \boldsymbol{e}_x$ and $\nabla^2 = \partial_x^2$. We consider now in turn the equations of our perturbation theory. By inserting the *ansatz* (5) into the LLG equation (3) using the effective field (4) we get two nontrivial equations to zeroth order,

$$\begin{aligned}
2\gamma K_\perp \sin\theta \cos\phi \sin\phi &= 0, \\
2\gamma \cos\theta \sin\theta (K + K_\perp \sin^2\phi) &= 0.
\end{aligned}$$

SCIENTIFIC **REPORTS** | 6:31800 | DOI: 10.1038/srep31800    3

These are the $\theta$- and the $\phi$- component, respectively. The zeroth order equations give the static magnetization direction. We see that one solution is to set both angles to zero, $\theta = 0$ and $\phi = 0$. This solution is unchanged by adding an integer multiple of $\pi$ to either angle.

*Spin-wave amplitudes.* As in any perturbation theory[43], the first and higher order equations vanish identically when the excitation parameter $h$ goes to zero. Thus, $h \to 0$ leaves us with the static problem. For first and higher orders of the perturbation theory the excitation parameter $h$ is merely a convenient accounting device and can be treated as $h \to 1$. To first order in $h$ we once more get two nontrivial equations corresponding to the $\theta$- and $\phi$-components,

$$\partial_t s_\theta = -\frac{2\gamma}{m}\left(A\partial_x^2 - K - K_\perp - \frac{m\alpha}{2\gamma}\partial_t\right)s_\phi, \tag{6}$$

$$\partial_t s_\phi = +\frac{2\gamma}{m}\left(A\partial_x^2 - K - \frac{m\alpha}{2\gamma}\partial_t\right)s_\theta. \tag{7}$$

To solve these equations we make the assumption that the time-dependence is purely harmonic, $s_{\theta/\phi}(x, t) = s_{\theta/\phi}(x) \times \exp(-i\omega t)$. Equations (6) and (7) can be recast into one single equation by introducing the auxiliary variable $\psi(x) = s_\theta(x) + ic s_\phi(x)$. We multiply equation (6) by $ic$ and subtract equation (7). By requiring that the coefficient of $s_\phi$ is $ic$ times that of $s_\theta$ we get a second order equation in $c$. Solving this equation we find that $c = -[\gamma K_\perp \pm (\gamma^2 K_\perp^2 + m^2\omega^2)^{1/2}]/m\omega$. Although both signs are allowed mathematically, we must choose the one that makes sense physically. By choosing the minus sign we get real wave numbers in the intermediate steps leading up to the dispersion relation in equation (10). Thus, keeping to this choice we find the equation

$$\frac{2\gamma A}{m}\partial_x^2 \psi = \frac{\gamma}{m}\left(2K + K_\perp - \gamma^{-1}\sqrt{\gamma^2 K_\perp^2 + m^2\omega^2} - im\alpha\omega/\gamma\right)\psi \tag{8}$$

for $\psi(x)$. This equation is solved by the *ansatz* that $\psi(x)$ is a damped harmonic, $\psi(x) = \rho \exp(ikx) \exp(-\eta x/\Gamma)$. Here, $\rho$ is the amplitude of the excited spin-waves at the spin-wave source; $\eta = \text{sgn}(x)$, and by choosing $k > 0$ we ensure that $\psi$ represents damped waves traveling away from a source at the origin. If we substitute our *ansatz* for $\psi$ back into equation (8) and separate the real and imaginary part we find expressions for the damping length[44]

$$\Gamma = \frac{4\gamma A k}{m\alpha\omega} \tag{9}$$

and for the dispersion relation[24]

$$\omega^2 = \frac{4\gamma^2}{m^2}(Ak^2 + K + K_\perp)(Ak^2 + K). \tag{10}$$

The dispersion relation we have obtained is the usual dispersion relation for an ordinary ferromagnet. We do not observe the linear shift reported by Mills and Dzyaloshinskii[13] because the spin-wave propagation direction is parallel to the direction of the static magnetization. We make further comments on the case where the wave vector, the applied electric field and the static magnetization are mutually orthogonal at the end of this section.

Having obtained a solution for $\psi$ we need a second condition to solve for the spin-wave components. A reasonable condition is that they should be real, which gives

$$s_\theta(x, t) = +\rho \cos(kx - \omega t) \exp(-\eta x/\Gamma), \tag{11}$$

$$s_\phi(x, t) = -\frac{\rho m\omega \sin(kx - \omega t) \exp(-\eta x/\Gamma)}{\gamma K_\perp - \sqrt{\gamma^2 K_\perp^2 + m^2\omega^2}}. \tag{12}$$

*Magnetization reorientation.* Only the $\theta$- and the $\phi$-components contribute nontrivial equations to second order in $h$,

$$\frac{2\gamma}{m}\left(A\frac{d^2}{dx^2} - K\right)\Delta m_\theta = 2\gamma\gamma_0(E_y s_\phi \partial_x s_\phi - E_z s_\theta \partial_x s_\phi) \tag{13}$$

$$\frac{2\gamma}{m}\left(A\frac{d^2}{dx^2} - K - K_\perp\right)\Delta m_\phi = 2\gamma\gamma_0(E_z s_\theta \partial_x s_\theta - E_y s_\phi \partial_x s_\theta) \tag{14}$$

As we have already obtained expressions for the spin-wave components from the first order equations, these are ordinary differential equations on $\Delta m_\theta$ and $\Delta m_\phi$. By assumption, these are time-independent and we remove the time-dependence from the right-hand side by averaging over one spin-wave period $2\pi/\omega$. The fact that the spin-waves are exponentially damped provides us with the boundary conditions $\lim_{x\to\pm\infty}\Delta m_{\theta/\phi}(x) = 0$. The resulting solutions are

$$\Delta m_y(x) = -\frac{\eta\rho^2 m^2\omega\Gamma\gamma_0\left[E_z(Ak^2 + K)\Gamma k + E_y m\omega/2\gamma\right]\exp(-2\eta x/\Gamma)}{4\gamma K_\perp(m^2\omega^2/4\gamma^2 K_\perp - Ak^2 - K)(4A - \Gamma^2 K)}, \tag{15}$$

$$\Delta m_z(x) = -\frac{\eta\rho^2 m^2\omega\Gamma\gamma_0\left[E_z(Ak^2 + K)/m\omega - E_y\Gamma k/2\gamma\right]\exp(-2\eta x/\Gamma)}{(Ak^2 + K)(4A - \Gamma^2(K + K_\perp))}, \tag{16}$$

where we have switched to Cartesian coordinates for clarity.

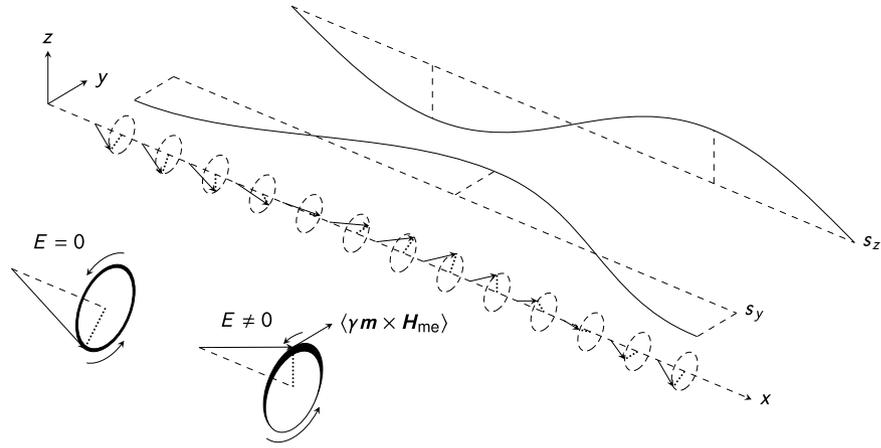

**Figure 4. A spin-wave in a homogeneously magnetized sample.** In the inset (down left) the weight of the stroke in the circle indicates the average time spent in that part of the circle over one period.

This demonstrates the existence of a magnon-mediated magnetoelectric torque acting on the homogeneous magnetization. Within the limits of validity of a perturbative theory, the magnitude of this torque scales linearly with the applied electric field. Its direction is controlled by the direction of the applied field. In the absence of a perpendicular anisotropy the relative magnitude of the two reorientation components is

$$\frac{\Delta m_y}{\Delta m_z} = -\frac{E_y + E_z\Gamma k}{E_y\Gamma k - E_z}$$

For relevant damping lengths and wave numbers $\Delta m_z$ dominates $\Delta m_y$ by a factor of ~$10^6$ when the electric field is applied in the $y$-direction and *vice versa* when the electric field is applied in the $z$-direction.

*Induced polarization.* The induced time-averaged electric polarization is directly proportional to the reorientation of the magnetization

$$\boldsymbol{P} = \frac{2\eta m\gamma_0}{\Gamma}\Delta\boldsymbol{m}. \tag{17}$$

However, while the reorientation of the magnetization changes sign at the origin (position of the source), the polarization does not due to the extra factor of $\eta = \text{sgn}(x)$.

Although the polarization is directly proportional to the magnetization reorientation, their behavior in the limit $\alpha \to 0$ differs due to the extra factor of $1/\Gamma$. In the limit where the damping is large ($\alpha \to \infty$ and $\Gamma \to 0$) all dynamics is quenched, and both $\Delta\boldsymbol{m}$ and $\boldsymbol{P}$ go to zero. However, in the limit where there is no damping ($\alpha \to 0$ and $\Gamma \to \infty$) the time-averaged polarization goes to zero, as pointed out by Mostovoy[8], but we still observe a finite magnetization reorientation.

*Simplified mechanism.* The simplest possible system that exhibits a magnon-mediated magnetoelectric torque is a homogeneously magnetized sample. The spin-wave propagation in this system is illustrated by the damping-less spin-chain in Fig. 4. Writing the magnetization of this chain as $\boldsymbol{m}(x, t) = m\boldsymbol{e}_x + s_y(x, t)\boldsymbol{e}_y + s_z(x, t)\boldsymbol{e}_z$ we get a contribution

$$\boldsymbol{H}_{\text{me}} = 2\gamma_0 E(\partial_x m\boldsymbol{e}_y - \partial_x s_y\boldsymbol{e}_x) = -2\gamma_0 E\partial_x s_y\boldsymbol{e}_x$$

| Parameter | Value | Unit |
|---|---|---|
| Gyromagnetic ratio, $\gamma$ | −26 | GHz/T |
| Magnetization, $m$ | 6 | kA/m |
| Exchange stiffness, $A$ | 5 | pJ/m |
| Easy axis anisotropy, $K$ | 0.5 | kJ/m$^3$ |
| Gilbert damping parameter, $\alpha$ | −0.05 | 1 |
| Inhomogeneous magnetoelectric coupling, $\gamma_0$ | 0.1 | psm/A |

**Table 1. Material constants used in the numerical solution of the LLG equation.** These values correspond to the iron garnets (BiR)$_3$(FeGa)$_5$O$_{12}$ (R = Lu, Tm) considered by Logginov et al.[34,35].

to the effective field from the inhomogeneous magnetoelectric effect when $\boldsymbol{E} = E\boldsymbol{e}_y$. If this contribution to the effective field had been constant on the time-scale of the spin-waves, it would only have changed the precession frequency. However, it is not. The torque exerted on the magnetization is

$$\gamma \boldsymbol{m} \times \boldsymbol{H}_{\mathrm{me}} = 2\gamma\gamma_0 E(s_y \partial_x s_y \boldsymbol{e}_z - s_z \partial_x s_y \boldsymbol{e}_y).$$

When averaging over one oscillation period the $z$-component vanishes; thus, the magnetization experiences a net torque in the positive $y$-direction ($\gamma < 0$),

$$\langle \gamma \boldsymbol{m} \times \boldsymbol{H}_{\mathrm{me}} \rangle = -2\gamma\gamma_0 E \langle s_z \partial_x s_y \rangle \boldsymbol{e}_y. \quad (18)$$

As illustrated in the inset of Fig. 4, this torque opposes the precession of the spin when it moves in the negative $y$-direction and boosts the precession of the spin when it moves in the positive $y$-direction. When the spins precess in the counterclockwise direction, $s_z > 0$ in the half-period where the spins are slowed by the magnetoelectric torque and $s_z < 0$ in the half-period where the spins are accelerated by the magnetoelectric torque. If we average over one period, the spins spend more time having $s_z > 0$ than $s_z < 0$, so the end result is that the system has acquired a net magnetic moment in the positive $z$-direction.

This mechanism can also be used to explain the results of Manchon et al.[30] and Linder[31], as is easily seen by repeating the calculation above using the effective fields employed in these papers. As such, it represents a unifying framework for magnon-mediated torques on the homogeneous magnetization. In particular, it identifies the following criterion for a system to exhibit similar magnon-mediated torques: That the effective field should contain a gradient of one of the transverse magnetization components.

Upon introducing the expansion (5) of the magnetization in $h$, we stated that the reorientations of the magnetization $\Delta m_{\theta/\phi}(\boldsymbol{r})$ are of second order in $h$ or higher since spin-waves in an ordinary ferromagnet do not induce such deviations. The fact that equation (18) is second order in the spin-wave amplitudes provides justification that $\Delta m_{\theta/\phi}(\boldsymbol{r})$ are in fact exactly of second order.

**Full numerical solution.** To get further insight into the properties of the magnon-mediated magnetoelectric torque we solve the LLG equation numerically using an adaptive centered implicit scheme for time and space discretization in Maple[45]. We use the effective field given in equation (4) and material constants corresponding to the iron garnets (BiR)$_3$(FeGa)$_5$O$_{12}$ (R = Lu, Tm) considered by Logginov et al.[34,35], see Table 1. We solve the system on a grid that is 8 μm long at grid-points spaced 4 nm apart. The system is initially homogeneously magnetized in the $x$-direction. Spin-waves are excited by applying a magnetic field $\boldsymbol{H}(t) = H_0 \sin(2\pi f t)\boldsymbol{e}_y$ to 24 grid-points at either side of the origin. The excitation amplitude is $H_0 = 0.2$ T and the excitation frequency is kept at 5 GHz. To avoid spin-wave reflection at the sample ends, we implement absorbing boundary conditions by increasing the Gilbert damping to $|\alpha| = 1$ inside 1 μm wide regions at either end of the sample[44]. The electric field is applied in the $y$-direction, $\boldsymbol{E} = E\boldsymbol{e}_y$, and we set the perpendicular anisotropy to zero.

Figure 5(a) illustrates the symmetry of the effect. These steady-state magnetization profiles have been obtained for an applied electric field of ±1.5 V/cm. As can also be seen from equation (16) the sign of the torque is determined by the propagation direction of the spin-waves, thus the resulting magnetization is antisymmetric about the spin-wave source. The torque also scales linearly with the applied electric field, so switching the sign of the field switches the sign of the magnetization reorientation.

Figure 5(b) through Fig. 5(d) shows a plots of the time-averaged analytical expression (16) versus the corresponding numerical profile for three different field values. The perturbation theory successfully predicts the magnitude and spatial decay of the torque. As can be seen from equation (16), the decay length of the magnetization reorientation is half that of the spin-waves as the reorientation is second order in $h$. The perturbative calculation predicts that the torque should scale linearly with the applied electric field. Plotting the magnetization shift at a fixed position ($x = 0.25$ μm) as a function of the applied electrical field we indeed observe a linear field-dependence, see Fig. 5(e).

**Comment on the linear shift of the dispersion.** Mills and Dzyaloshinskii[13] have pointed out that the inhomogeneous magnetoelectric effect will produce a shift of the spin-wave frequency with respect to that of an ordinary ferromagnet on the form

$$\Delta \omega = -C(\boldsymbol{E} \times \boldsymbol{M}_s) \cdot \boldsymbol{k}. \quad (19)$$

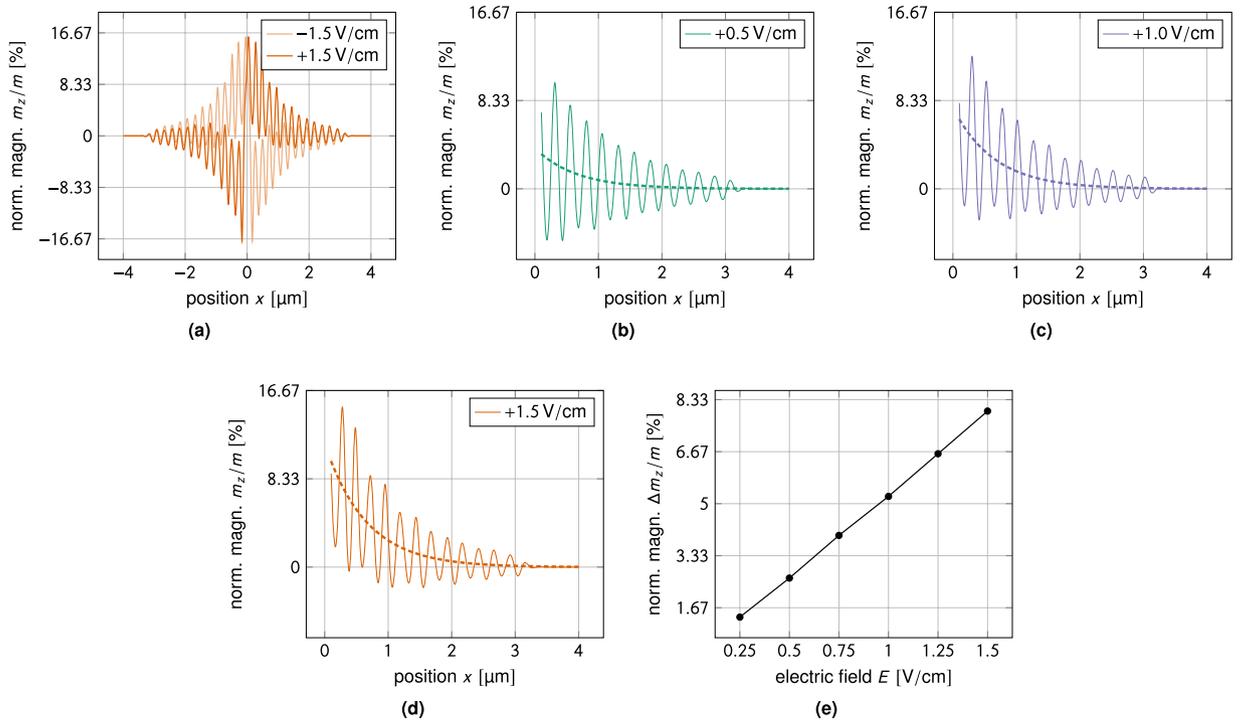

**Figure 5. Numerical results for magnon-mediated magnetoelectric torque.** When the electric field is applied in the $y$-direction the magnetization reorientation takes place in the $z$-direction. (**a**) Symmetry of the effect. Steady-state magnetization profiles obtained for $E = \pm 1.5$ V/cm. The sign of the torque is determined by the propagation direction of the spin-waves, thus the resulting magnetization is antisymmetric about the spin-wave source ($x = 0$). The torque also scales linearly with the applied electric field, so switching the sign of the field switches the sign of the magnetization reorientation. (**b**) through (**d**) Time-averaged analytical fit to the numerical profiles. The analytical theory successfully predicts the magnitude and spatial decay of the torque. The decay length of the spin-waves is twice that of the torque. (**e**) Linear dependence on applied field. Plotting the magnetization shift at a fixed position ($x = 0.25$ μm) as a function of the applied electrical field we observe a linear field-dependence.

Here $E$ is the applied electric field, $M_s$ is the saturation magnetization, $k$ is the spin-wave wave vector and $C$ is some positive constant. Liu and Vignale[14,15] have established in a microscopic calculation that yttrium iron garnet (YIG; $Y_3Fe_5O_{12}$) hosts the inhomogeneous magnetoelectric effect and that the phase shift should be observable. This experiment was carried out by Zhang et al.[16], showing good agreement between theory and experiment. In the notation of Liu and Vignale[15] the constant $C$ is written

$$C = \frac{|\gamma| eJ}{E_{so}},$$

where $\gamma$ is the gyromagnetic ratio, $e$ is the electron charge, $J = 1.6 \cdot 10^{-22}$ Jm/A$^2$ is the YIG exchange coupling and the energy $E_{so} = 4.8 \cdot 10^{-19}$ J is inversely proportional to the strength of the YIG spin-orbit coupling.

In the notation of Mills and Dzyaloshinskii[13], which is closer to our notation, the constant $C$ can be written $C = -|\gamma| b$, where $b$ is the inhomogeneous magnetoelectric coupling constant $b = b_1 + b_2$ used by Mills and Dzyaloshinskii. Thus, for YIG we can calculate $\gamma_0 = b_1 = b_2$ to be

$$\gamma_0 = \frac{eJ}{2E_{so}} = 2.7 \cdot 10^{-23} \text{ sm/A}.$$

While this value is large enough to produce a measurable spin-wave phase shift, it is far too small to produce a measurable reorientation of the time-averaged magnetization in YIG without exceeding the dielectric breakdown field. However, by turning this argument around, the difference of about ten orders of magnitude indicates that it should be possible to observe gigantic phase shifts if one were to redo the experiment of Zhang et al. using the iron garnets of Logginov et al.

Writing out the frequency shift as a triple product on the form of equation (19), as was done by Mills and Dzyaloshinskii[13], emphasizes the importance of the experimental geometry in order to observe this effect. Such geometrical considerations will also be of importance for observing the magnetization reorientation, as is easily seen by repeating the perturbative calculation leading up to equations (15) and (16) for perpendicular $m$ and $k$. In this geometry, which is the geometry of Liu and Vignale[15] and of Zhang et al.[16], no magnetization reorientation is observed.

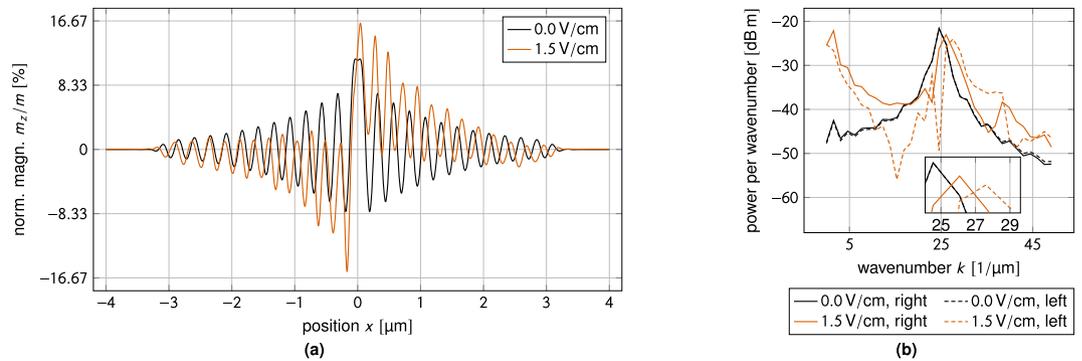

**Figure 6. Higher order phase shift due to the magnon-mediated magnetoelectric torque.** (**a**) Visibility of phase shift in magnetization profiles. Steady-state magnetization profiles obtained for $E = 1.5$ V/cm and in the absence of an applied electric field. A higher order correction of the ordinary ferromagnetic dispersion relation induces a phase shift when the electric field is applied. (**b**) Power spectral density of the magnetization profiles in (**a**) computed from the data to the right (left) of the source. In the presence of an applied electric field the power spectrum exhibits two peaks. The peak to the left is an artifact introduced by the magnetization reorientation. The one to the right—which corresponds to the spin-wave—is slightly displaced towards higher wavenumbers (smaller wavelengths) as compared to the zero field case.

Further insight can be gained by a closer inspection of the magnetization curves in Fig. 6(a). Here, an electric field-dependent phase shift can be observed which does not follow from the first order dispersion relation (10), but which is a higher order effect. This field-dependent difference in wavelength is easily recognized in the power spectrum shown in Fig. 6(b), which reveals an increase in wavenumber (decrease in wavelength) at $E = 1.5$ V/cm as compared to zero applied field.

**Proposals for experiments.** The effect we predict is dependent on a reasonable value for the inhomogeneous magnetoelectric coupling constant. One particular class of such materials are the rare earth manganites considered by Mostovoy[8], which have a spiral magnetization. Another class of materials which potentially hosts the inhomogeneous magnetoelectric effect is the iron garnets studied by Logginov and co-workers[34,35], which admit a homogeneous magnetization. We have used material values corresponding to these garnets in the numerical part of this work. The mechanism behind the experimental observations of Logginov et al. is currently debated[35,46]. Whether or not it turns out that these iron garnets host an inhomogeneous magnetoelectric effect is irrelevant to the main message in the present paper. We only use these compounds as an example.

In materials with the Dzyaloshinskii–Moriya interaction the crystal structure favors a canted spin structure[28,29]. Sergienko and Dagotto[3,47] have suggested how the inverse mechanism, a form of exchange striction, can give rise to the inhomogeneous magnetoelectric effect. This model is usually applied to large magnetic structures such as magnetic spirals and domain walls, which are on the energy scale of the demagnetization field. To carry this—or any other ionic displacement mechanism[48]—over to weaker magnetic inhomogeneities such as spin-waves is not trivial. It is not obvious that a coupling constant measured using a domain wall[34] should be applicable to spin-waves, although that is what we assume by using these values.

To detect the magnon-mediated magnetoelectric torque we propose to either measure the magnetization reorientation directly or to measure the induced polarization. As pointed out previously, the reorientation scales linearly as a function of the electric field and decays exponentially away from the source. Using the example values from Fig. 5(e), a reorientation on the order of 480 A/m should be unproblematic at a distance of 0.25 μm away from the source for an applied electric field of 1.5 V/cm. Polar MOKE (magneto-optic Kerr effect) techniques have previously been successfully applied to measure magnetization reorientations on the order of 56 A/m by Fan *et al.*[49] during their study of the spin-orbit torque. As pointed out in the previous section, the wave vector of the spin-waves must be parallel to the saturation magnetization to observe the magnetization reorientation due to the magnetoelectric torque. An experimental set-up must then use the less sensitive longitudinal MOKE[50], as illustrated in Fig. 7. However, the effect should still be within the reach of current experimental techniques.

The second possibility is to measure the induced polarization. When averaging over the period of oscillation of the spin-waves, we find that the time-averaged polarization is proportional to the induced magnetization shift, see equation (17). For the material values we have considered, a reorientation of 480 A/m corresponds to a time-averaged polarization on the order of 40 μC/cm$^2$. To measure the polarization, the MOKE laser in Fig. 7 would have to be replaced by localized electrodes on the top and on the bottom of the structure.

### Magnon-induced domain wall motion in a multiferroic

As shown in Fig. 2, we consider the same ferromagnetic wire as in the previous section, but assume now that it contains a Néel domain wall. (Bloch domain walls induce no electric polarization and are immune to magnetoelectric effects[7,8]).

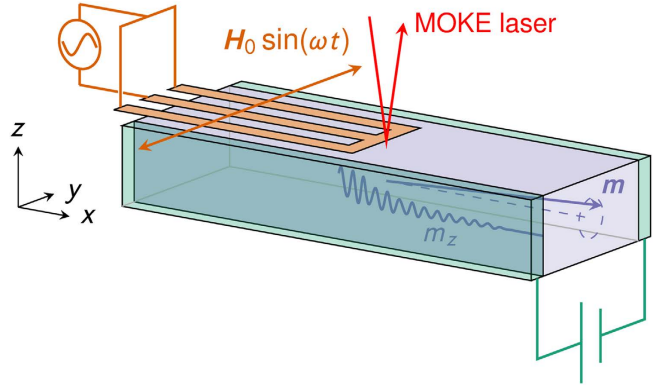

**Figure 7. Longitudinal MOKE measurement of the magnetization reorientation.** Spin-waves are injected using a coplanar waveguide as used in PSWS (propagating spin-wave spectroscopy[56,57]) and the electric field is generated by a parallel plate capacitor. The magnetization reorientation decays exponentially away from the spin-wave source, but scales linearly with the applied electric field.

**Perturbation theory.** We first perform analytical calculations for the magnon-induced domain wall dynamics using a perturbation theory, completely analogous to what we did for the homogeneously magnetized ferromagnet. However, this time we let the angles $\theta$ and $\phi$ from Fig. 3 be time- and position-dependent.

*Static magnetization.* Just as before, we obtain the equations of our perturbation theory by inserting the *ansatz* (5) into the LLG equation (3) using the effective field (4). To zeroth order in $h$ there is no time-dependence in the problem, and we obtain two nontrivial equations,

$$A\partial_x^2 \theta = \cos\theta \sin\theta (K + K_\perp \sin^2\phi), \tag{20}$$

$$\gamma_0 E_y (\partial_x \theta) \sin\phi - \gamma_0 E_z (\partial_x \theta) \cos\phi - \frac{2K_\perp}{m^2} \cos\phi \sin\phi = 0. \tag{21}$$

Equation (20) gives us the well known Walker domain wall profile[51] for $\theta$. Thus, the ground state of the system is a head-to-head or a tail-to-tail Néel wall. For $\phi = n\pi$, $n = 0, 1, 2, \ldots$, which minimizes the perpendicular anisotropy energy, equation (21) demands that $E_z = 0$, that is, no electric field component along the hard axis. For $K_\perp = 0$ equation (21) can be solved to give $\phi = n\pi + \arctan(E_z/E_y)$, $n = 0, 1, 2, \ldots$. We conclude that unless the electric field is applied perpendicular to the hard axis, the inhomogeneous magnetoelectric torque and the perpendicular anisotropy torque will compete. For simplicity, we set $K_\perp \equiv 0$ and $E_z \equiv 0$ in the remainder of this section.

*Spin-wave amplitudes.* Assuming that the domain wall position $X(t)$ is second order in $h$[42], we obtain two nontrivial equations to first order in $h$,

$$\partial_t s_\theta = -\frac{2\gamma}{m}\left( A\partial_x^2 + K[1 - 2\cos^2\theta] + \frac{(-1)^n \gamma_0 E_y m^2}{\lambda}\sin\theta - \frac{m\alpha}{2\gamma}\partial_t \right) s_\phi, \tag{22}$$

$$\partial_t s_\phi = +\frac{2\gamma}{m}\left( A\partial_x^2 + K[1 - 2\cos^2\theta] - \frac{m\alpha}{2\gamma}\partial_t \right) s_\theta. \tag{23}$$

We can get rid of the angle $\theta$ using the identities $\sin\theta = \text{sech}\,\xi$ and $\cos\theta = \tanh\xi$ which are valid for a Walker domain wall[52]. Here, $\xi = [x - X(t)]/\lambda$ and $\lambda = \sqrt{A/K}$ is the domain wall width. Using the same trick as before, we can recast these two equations into one equation by introducing the auxiliary variable $\psi(x) = s_\theta(x) + ics_\phi(x)$. However, this time we find that $c = \text{sech}\,\xi \left[(-1)^n \gamma\gamma_0 E_y m \pm (\gamma^2\gamma_0^2 E_y^2 m^2 + \lambda^2\omega^2 \cosh^2\xi)^{1/2}\right]/\lambda\omega$ and that

$$\frac{2\gamma K}{m}\partial_\xi^2 \psi = \frac{\gamma}{m}\left( 2K[1 - 2\,\text{sech}^2\xi] - \frac{(-1)^n \gamma_0 E_y m^2}{\lambda}\text{sech}\,\xi \right.$$
$$\left. \pm \frac{m}{\gamma\lambda}\text{sech}\,\xi\sqrt{\gamma^2\gamma_0^2 E_y^2 m^2 + \lambda^2\omega^2 \cosh^2\xi} \right)\psi. \tag{24}$$

This is a Schrödinger-like equation, but in the presence of an electric field the potential deviates from the reflectionless potential of Yan *et al.*[24]. The equation is even in $\xi$, so there is no dependence on the topological charge of the wall. However, the factor $(-1)^n$ ($n = 0, 1, 2, \ldots$) reveals that the potential is chirality-dependent. Chirality dependence was also observed for domain wall motion driven by a magnetic field by Chen *et al.*[32].

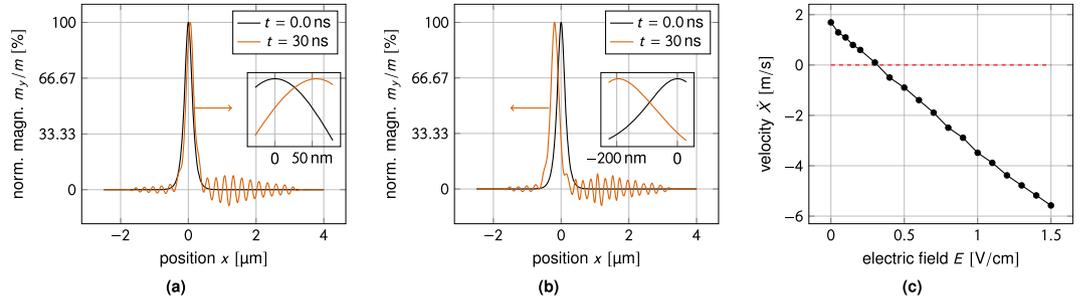

**Figure 8. Numerical results for electric field-controlled domain wall motion in a multiferroic.** The applied electric field can control both the direction and the velocity of the domain wall. (**a**) Zero applied field. In the absence of an applied electric field the domain wall moves towards the spin-wave source to conserve angular momentum. (**b**) Applied electric field. With an applied electric field the domain wall can be made to move away from the spin-wave source. (**c**) Linear dependence of the domain wall velocity on the applied electric field. The domain wall velocity is a linear function of the applied electric field. For some critical field the domain wall stops and its direction of motion is reversed.

**Full numerical solution.** The complicated potential in equation (24) makes it difficult to continue the analytical calculation. However, by solving the LLG equation numerically we can get further insight into the domain wall dynamics. We continue to use the iron garnet values from Table 1. We solve the system on an asymmetric grid that is 6.5 μm long at grid-points spaced 4 nm apart. The initial profile is a domain wall with positive topological charge ($\partial_x \theta > 0$) and positive chirality ($\phi = 0$) centered at the origin. Spin-waves are excited by applying a magnetic field $H(t) = H_0 \sin(2\pi f t) e_y$ to 24 grid-points at either side of $x = 1.2$ μm. The excitation amplitude is $H_0 = 0.2$ T and the excitation frequency is kept at 5 GHz. To avoid spin-wave reflection at the sample ends, we implement absorbing boundary conditions by increasing the Gilbert damping to $|\alpha| = 1$ inside 1 μm wide regions at either end of the sample[44]. The electric field is applied in the $y$-direction, $E = E e_y$, and the perpendicular anisotropy is set to zero.

Figure 8(a) is a reproduction of the well-known results of Yan *et al.*[24]. In the absence of an applied electric field the domain wall moves towards the spin-wave source in order to conserve angular momentum. Figure 8(b) shows that the magnon-mediated magnetoelectric torque can be used to reverse the velocity of the domain wall and make it travel away from the spin-wave source. The domain wall velocity is a linear function of the applied electric field, as shown in Fig. 8(c). At a critical applied electric field the domain wall stops and its direction of motion is reversed.

As suggested by Yan *et al.*[37], domain wall motion away from the spin-wave source is caused by linear momentum transfer, with linear momentum formally defined as the generator of magnetic translations[53,54]. To identify the momentum of a magnetization texture is not trivial, and an essential first step is to realize that the linear momentum of a ferromagnetic soliton is not directly related to its velocity, but to its configuration. This gives the conserved momentum of a domain wall some counter-intuitive properties—for instance, the momentum of a stationary domain wall can be nonzero[37,53]. Following Tchernyshyov[53] we calculate the conserved linear momenta of a magnetization texture consisting of circularly polarized spin-waves superposed on a Walker domain wall (axially symmetric system). The conserved linear momentum attributable to the domain wall and the spin-waves is, respectively

$$P_{\text{DW}} = C + 2m\phi/\gamma, \tag{25}$$

$$P_{\text{SW}} = C - \frac{\rho^2}{2\gamma m}\left(2\phi - \int_{-\infty}^{+\infty} k \, dx\right), \tag{26}$$

where $C$ denotes the momentum of the reference magnetization profile. Conservation of momentum, $0 = dP/dt = dP_{\text{DW}}/dt + dP_{\text{SW}}/dt$, allows us to solve for the rate of change of the azimuthal angle,

$$\dot{\phi} = \frac{\rho^2}{2(\rho^2 - 2m^2)} \int_{-\infty}^{+\infty} \dot{k} \, dx,$$

where the dot denotes a time-derivative. Thus, a change in the linear momentum (wavenumber) of the spin-waves generates domain wall rotation. This suggests that an effective description of the induced dynamics can be made in terms of an applied magnetic field[37,38]. In an axially symmetric system (no perpendicular anisotropy) the collective coordinate equations of a Walker domain wall subject to an applied magnetic field read[55],

$$(1 + \alpha^2)\dot{\phi} = -\gamma H, \tag{27}$$

$$(1 + \alpha^2)\frac{\dot{X}}{\lambda} = +\alpha\gamma H, \tag{28}$$

($\gamma < 0$, $\alpha < 0$). This gives

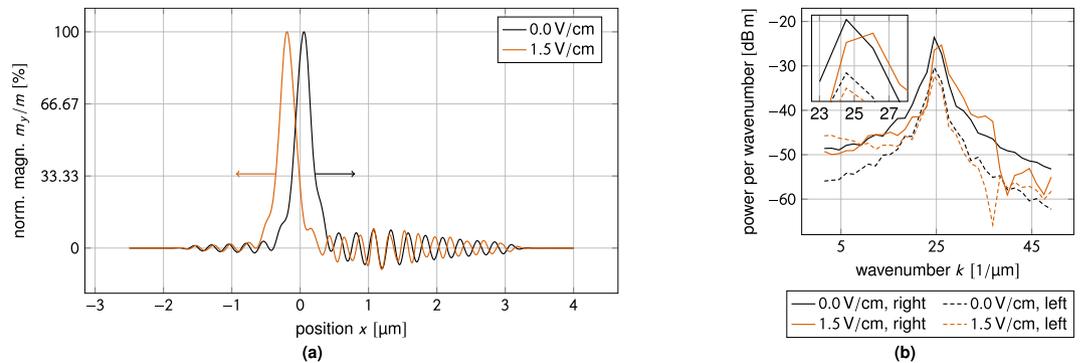

**Figure 9. Electric field-dependent linear momentum transfer to the domain wall.** (**a**) Phase shift difference before and after domain wall transmission. Steady-state magnetization profiles obtained for $E = 1.5$ V/cm and in the absence of an applied electric field. To the right of the wall (before transmission through the domain wall) there is a pronounced phase shift between the spin-waves in the two magnetization profiles. To the left of the wall (after domain wall transmission) the wavelengths of the spin-waves are practically identical. (**b**) Power spectral density of the magnetization profiles in (**a**) computed from the data to the right (left) of the source. The spurious peak in the power spectrum is now absent as there is no reorientation of the $y$-component of the magnetization. Whereas the position of the peak in the power spectrum is unchanged upon spin-wave transmission through the domain wall in the zero field case, the peak is shifted towards smaller wavenumbers in the presence of an applied electric field. This corresponds to a transfer of linear momentum to the domain wall.

$$H_{\text{eff}} = \frac{\rho^2(1+\alpha^2)}{2\gamma(2m^2-\rho^2)} \int_{-\infty}^{+\infty} k \mathrm{d}x. \tag{29}$$

The resulting domain wall velocity, $\dot{X}$, is small since it is proportional to the Gilbert damping. In the case of a domain wall driven by an applied magnetic field, the domain wall velocity can be increased drastically at small driving fields by breaking the axial symmetry of the system using a perpendicular anisotropy. Wang et al.[38] argued that the above effective description of linear momentum transfer can be carried over to this case and that an equation similar to equation (29) is valid even in the presence of a perpendicular anisotropy. Using this description they were able to explain domain wall motion driven by linear momentum transfer from spin-waves in a Dzyaloshinskii–Moriya ferromagnet.

In the present case, the axial symmetry of the system is broken by the applied electric field. Since we have not been able to solve equation (24) for the spin-wave amplitudes, we are not able to repeat the analysis leading up to equation (29) in the presence of the inhomogeneous magnetoelectric effect. However, we argue that—just as for the Dzyaloshinskii–Moriya ferromagnet—the underlying physics is the same and that linear momentum transfer from the spin-waves leads to domain wall motion that can be described using an effective Zeeman field.

The mechanism of momentum transfer from the spin-waves to the domain wall is somewhat similar to the one suggested by Wang et al.[38] for linear momentum transfer in the presence of the Dzyaloshinskii–Moriya interaction. As can be seen in Fig. 9(a) the application of an electric field does not cause significant spin-wave reflection off the domain wall. However, in the presence of an electric field the wavelength changes upon spin-wave transmission through the domain wall. This is captured by the power spectrum in Fig. 9(b). The reduction of the spin-wave wavenumber upon transmission implies that the transmitted spin-waves carry less linear momentum. This change in linear momentum is absorbed by the domain wall.

Since the effective Zeeman field is linear in the momentum transfer[37,38] and the domain wall velocity is a linear function of the applied magnetic field below Walker breakdown[32,51], the linear dependence of the domain wall velocity on the electric field in Fig. 8(c) must be taken as a sign that the momentum transfer is linear in the applied electric field, just as it is linear in the DMI constant[38]. This is not too surprising, given the linear dependence on the electric field found in equation (24). The dependence of the momentum transfer on the electric field can also be observed in a sequence of power spectra like the one in Fig. 9(b) taken at increasing electric fields. As the electric field is increased, the wavenumber of the spin-waves to the right of the domain wall is shifted away from the zero field peak, which is just what we pointed out for the homogeneously magnetized case in Fig. 6(b).

## Conclusion

To conclude, we have demonstrated analytically and numerically that the inhomogeneous magnetoelectric effect induces a magnon-mediated reorientation of a homogeneous magnetization, and we have provided an explanation of the mechanism behind this effect. This reorientation is not fixed by material constants like the ones discovered for Dzyaloshinskii–Moriya ferromagnets by Manchon et al.[30] and for topological insulator–ferromagnet heterostructures by Linder[31], but is tunable by the applied electric field. Its magnitude increases linearly with the electric field and an effect on the order of 8% of the saturation magnetization should not be problematic. This is almost an order of magnitude larger than the reorientation reported by Manchon et al. for Dzyaloshinskii–Moriya ferromagnets.

We have also shown that the sign and magnitude of the velocity of a magnon-driven domain wall can be controlled by the applied electric field in the presence of the inhomogeneous magnetoelectric interaction. Domain wall motion towards the spin-wave source is due to angular momentum conservation while domain wall motion away from the source is due to linear momentum transfer. The mechanism of linear momentum transfer is quite similar to the mechanism suggested by Wang et al.[38] for Dzyaloshinskii–Moriya ferromagnets, and the domain wall velocity scales linearly with the electric field.

## References

1. Eerenstein, W., Mathur, N. D. & Scott, J. F. Multiferroic and magnetoelectric materials. *Nature* **442**, 759–765, URL http://www.nature.com/doifinder/10.1038/nature05023 (2006).
2. Spaldin, N. A. The renaissance of magnetoelectric multiferroics. *Science* **309**, 391–392, URL http://www.sciencemag.org/cgi/doi/10.1126/science.1113357 (2005).
3. Cheong, S.-W. & Mostovoy, M. Multiferroics: A magnetic twist for ferroelectricity. *Nat. Mater.* **6**, 13–20, URL http://www.nature.com/doifinder/10.1038/nmat1804 (2007).
4. Dong, S., Liu, J.-M., Cheong, S.-W. & Ren, Z. Multiferroic materials and magnetoelectric physics: Symmetry, entanglement, excitation, and topology. *Adv. Phys.* **64**, 519–626, URL http://www.tandfonline.com/doi/full/10.1080/00018732.2015.1114338 (2015).
5. Dzyaloshinskii, I. E. On the magneto-electrical effect in antiferromagnets. *J. Exp. Theor. Phys.* **10**, 628–629, URL http://www.jetp.ac.ru/cgi-bin/e/index/e/10/3/p628?a=list (1960).
6. Schmid, H. Some symmetry aspects of ferroics and single phase multiferroics. *J. Phys. Condens. Mat.* **20**, 434201, URL http://stacks.iop.org/0953-8984/20/i=43/a=434201?key=crossref.13df1153dc47706416fa3f17637a8224 (2008).
7. Bar'yakhtar, V. G., L'vov, V. A. & Yablonskii, D. A. Inhomogeneous magnetoelectric effect. *JETP Lett.* **37**, 673–675, URL http://www.jetpletters.ac.ru/ps/1499/article_22895.shtml (1983).
8. Mostovoy, M. Ferroelectricity in spiral magnets. *Phys. Rev. Lett.* **96**, 067601, URL http://link.aps.org/doi/10.1103/PhysRevLett.96.067601 (2006).
9. Cano, A. & Kats, E. I. Electromagnon excitations in modulated multiferroics. *Phys. Rev. B* **78**, 012104, URL http://link.aps.org/doi/10.1103/PhysRevB.78.012104 (2008).
10. Tewari, S., Zhang, C., Toner, J. & Das Sarma, S. Goldstone modes and electromagnon fluctuations in the conical cycloid state of a multiferroic. *Phys. Rev. B* **78**, 144427, URL http://link.aps.org/doi/10.1103/PhysRevB.78.144427 (2008).
11. Cano, A. Theory of electromagnon resonances in the optical response of spiral magnets. *Phys. Rev. B* **80**, 180416, URL http://link.aps.org/doi/10.1103/PhysRevB.80.180416 (2009).
12. Zvezdin, A. K. & Mukhin, A. A. On the effect of inhomogeneous magnetoelectric (flexomagnetoelectric) interaction on the spectrum and properties of magnons in multiferroics. *JETP Lett.* **89**, 328–332, URL http://link.springer.com/10.1134/S0021364009070042 (2009).
13. Mills, D. L. & Dzyaloshinskii, I. E. Influence of electric fields on spin waves in simple ferromagnets: Role of the flexoelectric interaction. *Phys. Rev. B* **78**, 184422, URL http://link.aps.org/doi/10.1103/PhysRevB.78.184422 (2008).
14. Liu, T. & Vignale, G. Electric control of spin currents and spin-wave logic. *Phys. Rev. Lett.* **106**, 247203, URL http://link.aps.org/doi/10.1103/PhysRevLett.106.247203 (2011).
15. Liu, T. & Vignale, G. Flexoelectric phase shifter for spin waves. *J. Appl. Phys.* **111**, 083907, URL http://scitation.aip.org/content/aip/journal/jap/111/8/10.1063/1.4703925 (2012).
16. Zhang, X., Liu, T., Flatté, M. E. & Tang, H. X. Electric-field coupling to spin waves in a centrosymmetric ferrite. *Phys. Rev. Lett.* **113**, 037202, URL http://link.aps.org/doi/10.1103/PhysRevLett.113.037202 (2014).
17. Chumak, A. V., Serga, A. A. & Hillebrands, B. Magnon transistor for all-magnon data processing. *Nat. Commun.* **5**, 4700, URL http://www.nature.com/doifinder/10.1038/ncomms5700 (2014).
18. Chumak, A. V., Vasyuchka, V. I., Serga, A. A. & Hillebrands, B. Magnon spintronics. *Nat. Phys.* **11**, 453–461, URL http://www.nature.com/doifinder/10.1038/nphys3347 (2015).
19. Serga, A. A., Chumak, A. V. & Hillebrands, B. YIG magnonics. *J. Phys. D Appl. Phys.* **43**, 264002, URL http://stacks.iop.org/0022-3727/43/i=26/a=264002?key=crossref.6a073bfc1c8ba106fa1ba87bbfe37df3 (2010).
20. Khitun, A., Bao, M. & Wang, K. L. Magnonic logic circuits. *J. Phys. D Appl. Phys.* **43**, 264005, URL http://stacks.iop.org/0022-3727/43/i=26/a=264005?key=crossref.f9dd566068384c0f843c64d4110dbb4d (2010).
21. Khitun, A. & Wang, K. L. Non-volatile magnonic logic circuits engineering. *J. Appl. Phys.* **110**, 034306, URL http://scitation.aip.org/content/aip/journal/jap/110/3/10.1063/1.3609062 (2011).
22. Klingler, S. *et al.* Design of a spin-wave majority gate employing mode selection. *Appl. Phys. Lett.* **105**, 152410, URL http://scitation.aip.org/content/aip/journal/apl/105/15/10.1063/1.4898042 (2014).
23. Khitun, A. Multi-frequency magnonic logic circuits for parallel data processing. *J. Appl. Phys.* **111**, 054307, URL http://scitation.aip.org/content/aip/journal/jap/111/5/10.1063/1.3689011 (2012).
24. Yan, P., Wang, X. S. & Wang, X. R. All-magnonic spin-transfer torque and domain wall propagation. *Phys. Rev. Lett.* **107**, 177207, URL http://link.aps.org/doi/10.1103/PhysRevLett.107.177207 (2011).
25. Urazuka, Y., Imamura, K., Oyabu, S., Tanaka, T. & Matsuyama, K. Successive logic-in-memory operation in spin wave-based devices with domain wall data coding scheme. *IEEE Trans. Magn.* **50**, 1–3, URL http://ieeexplore.ieee.org/lpdocs/epic03/wrapper.htm?arnumber=6971479 (2014).
26. Fert, A., Cros, V. & Sampaio, J. Skyrmions on the track. *Nat. Nanotechnol.* **8**, 152–156, URL http://www.nature.com/doifinder/10.1038/nnano.2013.29 (2013).
27. Schütte, C. & Garst, M. Magnon-skyrmion scattering in chiral magnets. *Phys. Rev. B* **90**, 094423, URL http://link.aps.org/doi/10.1103/PhysRevB.90.094423 (2014).
28. Dzyaloshinskii, I. E. A thermodynamic theory of "weak" ferromagnetism of antiferromagnetics. *J. Phys. Chem. Solids* **4**, 241–255, URL http://linkinghub.elsevier.com/retrieve/pii/0022369758900763 (1958).
29. Moriya, T. Anisotropic superexchange interaction and weak ferromagnetism. *Phys. Rev.* **120**, 91–98, URL http://link.aps.org/doi/10.1103/PhysRev.120.91 (1960).
30. Manchon, A., Ndiaye, P. B., Moon, J.-H., Lee, H.-W. & Lee, K.-J. Magnon-mediated Dzyaloshinskii-Moriya torque in homogeneous ferromagnets. *Phys. Rev. B* **90**, 224403, URL http://link.aps.org/doi/10.1103/PhysRevB.90.224403 (2014).
31. Linder, J. Improved domain-wall dynamics and magnonic torques using topological insulators. *Phys. Rev. B* **90**, 041412, URL http://link.aps.org/doi/10.1103/PhysRevB.90.041412 (2014).
32. Chen, H.-B., Liu, Y.-H. & Li, Y.-Q. Electric field control of multiferroic domain wall motion. *J. Appl. Phys.* **115**, 133913, URL http://scitation.aip.org/content/aip/journal/jap/115/13/10.1063/1.4870711 (2014).
33. Dzyaloshinskii, I. E. Magnetoelectricity in ferromagnets. *Europhys. Lett.* **83**, 67001, URL http://stacks.iop.org/0295-5075/83/i=6/a=67001?key=crossref.4c3468a7a6c2ca649e3b02617a3ffd64 (2008).
34. Logginov, A. S., Meshkov, G. A., Nikolaev, A. V. & Pyatakov, A. P. Magnetoelectric control of domain walls in a ferrite garnet film. *JETP Lett.* **86**, 115–118, URL http://link.springer.com/10.1134/S0021364007140093 (2007).

35. Pyatakov, A. P. *et al.* Micromagnetism and topological defects in magnetoelectric media. *Phys.-Usp.* **58,** 981–992, URL http://stacks.iop.org/1063-7869/58/i=10/a=981?key=crossref.62c38296ce954ae637088d70bb2c471d (2015).
36. Mikhailov, A. V. & Yaremchuk, A. I. Forced motion of a domain wall in the field of a spin wave. *JETP Lett.* **39,** 354–357, URL http://www.jetpletters.ac.ru/ps/1299/article_19617.shtml (1984).
37. Yan, P., Kamra, A., Cao, Y. & Bauer, G. E. W. Angular and linear momentum of excited ferromagnets. *Phys. Rev. B* **88,** 144413, URL http://link.aps.org/doi/10.1103/PhysRevB.88.144413 (2013).
38. Wang, W. *et al.* Magnon-driven domain-wall motion with the Dzyaloshinskii-Moriya interaction. *Phys. Rev. Lett.* **114,** 087203, URL http://link.aps.org/doi/10.1103/PhysRevLett.114.087203 (2015).
39. Landau, L. D. & Lifshitz, E. M. On the theory of the dispersion of magnetic permeability in ferromagnetic bodies. *Phys. Zeitsch. der Sow.* **8,** 153–169 (1935).
40. Landau, L. D. & Lifshitz, E. M. On the theory of the dispersion of magnetic permeability in ferromagnetic bodies. *Ukr. J. Phys.* **53,** 14–22, URL http://ujp.bitp.kiev.ua/files/journals/53/si/53SI06p.pdf (2008).
41. Gilbert, T. A phenomenological theory of damping in ferromagnetic materials. *IEEE Trans. Magn.* **40,** 3443–3449, URL http://ieeexplore.ieee.org/lpdocs/epic03/wrapper.htm?arnumber=1353448 (2004).
42. Tveten, E. G., Qaiumzadeh, A. & Brataas, A. Antiferromagnetic domain wall motion induced by spin waves. *Phys. Rev. Lett.* **112,** 147204, URL http://link.aps.org/doi/10.1103/PhysRevLett.112.147204 (2014).
43. Griffiths, D. J. *Introduction to Quantum Mechanics*, 2 edn (Pearson, 2013).
44. Seo, S.-M., Lee, K.-J., Yang, H. & Ono, T. Current-induced control of spin-wave attenuation. *Phys. Rev. Lett.* **102,** 147202, URL http://link.aps.org/doi/10.1103/PhysRevLett.102.147202 (2009).
45. Waterloo Maple, *pdsolve*. Available at: http://www.maplesoft.com/support/help/Maple/view.aspx?path=pdsolve (Accessed: May 6, 2016) (2016).
46. Kabychenkov, A. F., Lisovskii, F. V. & Mansvetova, E. G. Magnetoelectric effect in garnet films with the induced magnetic anisotropy in a nonuniform electric field. *JETP Lett.* **97,** 265–269, URL http://link.springer.com/10.1134/S0021364013050081 (2013).
47. Sergienko, I. A. & Dagotto, E. Role of the Dzyaloshinskii-Moriya interaction in multiferroic perovskites. *Phys. Rev. B* **73,** 094434, URL http://link.aps.org/doi/10.1103/PhysRevB.73.094434 (2006).
48. Barone, P. & Picozzi, S. Mechanisms and origin of multiferroicity. *C. R. Phys.* **16,** 143–152, URL http://linkinghub.elsevier.com/retrieve/pii/S1631070515000109 (2015).
49. Fan, X. *et al.* Quantifying interface and bulk contributions to spin–orbit torque in magnetic bilayers. *Nat. Commun.* **5,** 3042, URL http://www.nature.com/doifinder/10.1038/ncomms4042 (2014).
50. Jiles, D. *Introduction to Magnetism and Magnetic Materials*, 1 edn. (Springer-Science+Business Media, 1991).
51. Schryer, N. L. & Walker, L. R. The motion of 180° domain walls in uniform dc magnetic fields. *J. Appl. Phys.* **45,** 5406–5421, URL http://scitation.aip.org/content/aip/journal/jap/45/12/10.1063/1.1663252 (1974).
52. Tatara, G., Kohno, H. & Shibata, J. Microscopic approach to current-driven domain wall dynamics. *Phys. Rep.* **468,** 213–301, URL http://linkinghub.elsevier.com/retrieve/pii/S0370157308002597 (2008).
53. Tchernyshyov, O. Conserved momenta of a ferromagnetic soliton. *Ann. Phys.* **363,** 98–113, URL http://linkinghub.elsevier.com/retrieve/pii/S0003491615003395 (2015).
54. Zak, J. Magnetic translation group. *Phys. Rev.* **134,** A1602–A1606, URL http://link.aps.org/doi/10.1103/PhysRev.134.A1602 (1964).
55. Shibata, J., Tatara, G. & Kohno, H. A brief review of field- and current-driven domain-wall motion. *J. Phys. D. Appl. Phys.* **44,** 384004, URL http://stacks.iop.org/0022-3727/44/i=38/a=384004?key=crossref.030ab431503338459b13bb288cf61a54 (2011).
56. Bailleul, M., Olligs, D. & Fermon, C. Propagating spin wave spectroscopy in a permalloy film: A quantitative analysis. *Appl. Phys. Lett.* **83,** 972, URL http://scitation.aip.org/content/aip/journal/apl/83/5/10.1063/1.1597745 (2003).
57. Sekiguchi, K. *et al.* Nonreciprocal emission of spin-wave packet in FeNi film. *Appl. Phys. Lett.* **97,** 022508, URL http://scitation.aip.org/content/aip/journal/apl/97/2/10.1063/1.3464569 (2010).

## Acknowledgements

We thank M. Mostovoy and T. Liu for helpful comments, and acknowledge support from the Outstanding Academic Fellows programme at NTNU and the Norwegian Research Council Grant No. 205591, No. 216700 and No. 240806.

## Author Contributions

J.L. and I.K. conceived the study. I.K. carried out the preliminary investigation. V.R. substantially expanded and shaped the project and wrote the manuscript. All authors reviewed the manuscript.

## Additional Information

**Competing financial interests:** The authors declare no competing financial interests.

**How to cite this article**: Risinggård, V. *et al.* Electric field control of magnon-induced magnetization dynamics in multiferroics. *Sci. Rep.* **6,** 31800; doi: 10.1038/srep31800 (2016).